\documentclass[article,aps,twocolumn,showpacs,amsmath,amssymb,superscriptaddress,nofootinbib,floatfix]{revtex4-2}

\usepackage{graphicx}% Include figure files
\usepackage{dcolumn}% Align table columns on decimal point
\usepackage{bm}% bold math
\usepackage[colorlinks=true,linkcolor=blue,citecolor=blue,urlcolor=blue]{hyperref}

\usepackage{array,booktabs,tabularx}
\usepackage[caption=false]{subfig}
\usepackage[USenglish]{babel}
\usepackage{braket}
\usepackage{xcolor}
\usepackage{amsfonts}
\usepackage{amssymb}
\usepackage[normalem]{ulem}
\usepackage{amsmath}

\begin{document}
\renewcommand{\figurename}{FIG.}

 \title{Incoherent Spectral Weight Emerging from a Van Hove Singularity in a Kagome Metal}

\author{C.-y. Lim}
\email{chanyoung.lim@dipc.org}
\affiliation{Donostia International Physics Center (DIPC), San Sebastián 20018, Spain}

\author{J. Deng}
\affiliation{Department of Applied Physics, Aalto University School of Science, Aalto 00076, Finland}

\author{A. Korshunov}
\affiliation{Donostia International Physics Center (DIPC), San Sebastián 20018, Spain}

\author{A. Kar}
\affiliation{Donostia International Physics Center (DIPC), San Sebastián 20018, Spain}

\author{D. Subires}
\affiliation{Donostia International Physics Center (DIPC), San Sebastián 20018, Spain}
\affiliation{University of the Basque Country (UPV/EHU), Bilbao 48080, Spain}

\author{H. Li}
\affiliation{Donostia International Physics Center (DIPC), San Sebastián 20018, Spain}

\author{Y. Jiang}
\affiliation{Donostia International Physics Center (DIPC), San Sebastián 20018, Spain}

\author{H. Hu}
\affiliation{Department of Physics, University of Science and Technology of China, Hefei, Anhui 230026, China}

\author{E. Modin}
\affiliation{CIC nanoGUNE, BRTA, San Sebastián 20018, Spain}

\author{P. T\"{o}rm\"{a}}
\affiliation{Department of Applied Physics, Aalto University School of Science, Aalto 00076, Finland}

\author{A. Kumar-Sharma}
\affiliation{Max Planck Institute for Chemical Physics of Solids, Dresden 01187, Germany}

\author{C. Shekhar}
\affiliation{Max Planck Institute for Chemical Physics of Solids, Dresden 01187, Germany}

\author{A. Louat}
\affiliation{Diamond Light Source Ltd, Harwell Science and Innovation Campus, Didcot OX11 0DE, UK}

\author{T. K. Kim}
\affiliation{Diamond Light Source Ltd, Harwell Science and Innovation Campus, Didcot OX11 0DE, UK}

\author{C. Felser}
\affiliation{Max Planck Institute for Chemical Physics of Solids, Dresden 01187, Germany}

\author{B. Andrei Bernevig}
\affiliation{Donostia International Physics Center (DIPC), San Sebastián 20018, Spain}
\affiliation{Department of Physics, Princeton University, Princeton, New Jersey 08544, USA}
\affiliation{IKERBASQUE, Basque Foundation for Science, Bilbao 48013, Spain}

\author{S. Blanco-Canosa}
\email{sblanco@dipc.org}
\affiliation{Donostia International Physics Center (DIPC), San Sebastián 20018, Spain}
\affiliation{IKERBASQUE, Basque Foundation for Science, Bilbao 48013, Spain}

\date{\today}

\begin{abstract}

Waterfall-like spectral features observed by angle-resolved photoemission spectroscopy (ARPES) are commonly associated with strong electronic correlations or electron-boson coupling. Here, we investigate this connection in the kagome metal LuFe$_6$Ge$_6$, a structurally simple member of the FeGe-derived family, using polarization-dependent ARPES, density functional theory and dynamical mean-field theory. We observe a pronounced vertical spectral feature at the \textit{L} point extending over several hundred meV in binding energy. Its orbital character is identified as predominantly Fe \textit{d}$_{yz}$, and its energy onset coincides with a van Hove singularity located approximately 0.3 eV below \textit{E}$_F$. Despite the waterfall-like spectral response, dynamical mean-field theory (DMFT) reveals only weak electronic correlations, while the calculated electronic structure closely reproduces the experimental bulk bands. We propose that the anomalous spectral weight may arise from inelastic scattering of the outgoing photoelectrons, potentially associated with the large density of states near the van Hove singularity. Our results suggest that waterfall-like features can emerge without strong electronic correlations and highlight the possible role of the underlying band structure and photoelectron energy-loss processes in their interpretation.
\end{abstract}

\maketitle

\date{\today}

One of the most characteristic manifestations of electronic correlations in a material is the modification of the electronic self-energy, \(\Sigma(k,\omega)\), which renormalizes both the dispersion and the linewidth of the bare electronic bands \cite{Morosan_2012,Matsuyama_2017}. In particular, electronic interactions can reduce the quasiparticle weight near \textit{E}$_F$, enhance the scattering rate, and generate incoherent spectral weight extending over a broad energy range. Such anomalous spectral features have been widely observed by angle-resolved photoemission spectroscopy (ARPES) in correlated materials \cite{Damascelli_2003,Sobota_2021}. A prominent example is the nearly vertical spectral feature associated with an abrupt change in the apparent quasiparticle velocity and lifetime. This high-binding-energy, incoherent spectral weight, commonly referred to as the ARPES “waterfall”, was observed in cuprate high-\(T_c\) superconductors \cite{PhysRevLett.91.167002, Graf_2007,PhysRevLett.93.117003, DSInosov_PhysRevLett.99.237002, PhysRevB.79.020504, Chang2013-yo}, and subsequently observed in superconducting iron pnictides \cite{Richard_2009,Koitzsch_2009}, ruthenates \cite{Iwasawa_2012}, nickelates \cite{10.1093/nsr/nwae194, doi:10.1126/sciadv.adr5116,Si_2024,XDu_Hund_waterfall}, and transition-metal chalcogenides \cite{Jang_2021}. Its widespread observation has been supported by dynamical mean-field theory (DMFT) calculations of Hubbard-type models, which have associated such incoherent spectral weight with Coulomb interactions \cite{Chang2024-ig}, coupling to collective bosonic excitations, coherent polarons \cite{Wang_2016,Verdi_2017,Li_2018}, spin-orbit coupling \cite{Liu_2015} and Hund's coupling \cite{Chang2024-ig,XDu_Hund_waterfall}. However, the microscopic origin of waterfall-like features remains debated: their appearance in ARPES does not by itself uniquely establish strong electronic correlations \cite{DSInosov_PhysRevLett.99.237002}, and similar spectral anomalies may arise from different sources of momentum- and energy-dependent scattering. This ambiguity highlights the need to establish a direct connection between the spectral feature and the underlying electronic structure.

Kagome materials provide a particularly attractive platform for investigating such spectroscopic signatures of electronic correlations. Their characteristic band structure naturally hosts flat bands (FBs), van Hove singularities (vHs), and Dirac crossings (DCs) \cite{Hu2022-xq,YHu2023_AV3Sb5}. In particular, flat bands and saddle-point vHs can give rise to an enhanced density of states (DOS). These features can enhance the effects of electronic interactions and promote interaction-driven instabilities such as charge density wave (CDW) \cite{Jiang2021-jq, PhysRevLett.125.247002,PhysRevLett.129.216402,Korshunov2023-nt,Teng2022-pf, PhysRevResearch.6.033222} and magnetic orders \cite{doi:10.1126/sciadv.abe2680,Thomas2025-jy,Nakatsuji2015-pf, PhysRevMaterials.7.064201, Ortiz2023-qn}. Spectroscopic signatures attributed to electronic correlations have been reported in kagome metals such as {FeGe} \cite{Teng2023-og}, {CsCr$_3$Sb$_5$} \cite{Li2025-ez} and {CsTi$_3$Bi$_5$} \cite{Yang2024-nz,Bigi_2025,Yang2023-tq}.

%Kagome materials provide a particularly attractive platform for investigating such spectroscopic signatures of electronic correlations. Their characteristic band structure naturally hosts regions of enhanced density of states (DOS), including flat bands (FBs), van Hove singularities (vHs), and Dirac crossings (DCs) \cite{Hu2022-xq,YHu2023_AV3Sb5}. These features can enhance the effects of electronic interactions and promote a variety of interaction-driven instabilities including charge-density-wave (CDW) \cite{Jiang2021-jq, PhysRevLett.125.247002,PhysRevLett.129.216402,Korshunov2023-nt,Teng2022-pf, PhysRevResearch.6.033222}, magnetic order \cite{doi:10.1126/sciadv.abe2680,Thomas2025-jy,Nakatsuji2015-pf, PhysRevMaterials.7.064201, Ortiz2023-qn}, superconductivity \cite{PhysRevLett.125.247002,Yang2024-nz, PhysRevB.94.094523, Gui2022-qd}, and topological phases \cite{Yang_2017, doi:10.1126/science.aav2873, Yin2020-wn, PhysRevB.104.235139}. Spectroscopic signatures attributed to electronic correlations have been reported in kagome metals such as FeGe \cite{Teng2023-og}, CsCr$_3$Sb$_5$ \cite{Li2025-ez} and CsTi$_3$Bi$_5$ \cite{Yang2024-nz,Bigi_2025,Yang2023-tq}.

The case of the kagome FeGe is particularly compelling \cite{Teng2022-pf} because DFT+DMFT calculations indicate that its Fe 3$d$ electrons exhibit pronounced signatures of Hund's coupling, including orbital differentiation and spin-orbital separation \cite{Huang_PRB_2020}. Its interplay with charge and magnetic order further complicates the identification of the microscopic origin of correlation-driven spectroscopic features. At the same time, the relatively simple FeGe structure can serve as a building block for a broader family of compounds, the “LEGO” architecture of FeGe-based materials \cite{Jiang_PRB_2025}. This family therefore provides an opportunity to systematically investigate how modifications of the structural environment influence the electronic spectrum and the signatures of electronic, charge, and spin correlations.

Here, we present a comprehensive investigation of the electronic structure of {LuFe$_6$Ge$_6$} using polarization-dependent ARPES, density functional theory (DFT), and DMFT. The combination of ARPES and theoretical calculations establishes the bulk band structure and orbital character of the observed electronic states. Beyond the well-defined dispersive bands, ARPES reveals a pronounced vertical spectral feature at the {L} point vHs extending over several hundred meV in binding energy. Comparison with DFT and DMFT distinguishes this anomalous spectral response from the underlying single-particle band structure and strong correlation-induced renormalization. These findings highlight that waterfall-like ARPES features do not necessarily constitute evidence of strong many-body correlations and that photoemission-specific scattering processes should also be considered in their interpretation.

\begin{figure}
    \centering
    \includegraphics{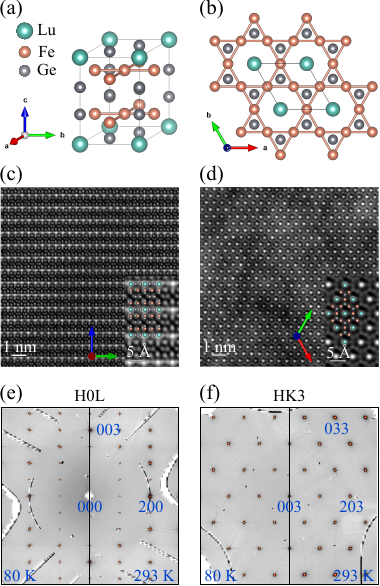}
    \caption{(a) Unit cell and (b) In-plane lattice structure of the kagome metal {LuFe$_6$Ge$_6$}. (c,d) Scanning transmission electron microscopy (STEM) images projected onto the (c) $bc$-plane and (d) $ab$-plne, respectively. The inset images in the bottom-right corners show magnified views overlaid with the atomic structure. (e,f) Comparison of the diffuse scattering results of {LuFe$_6$Ge$_6$} in the (e) {H0L} and (f) {HK3} planes, respectively, at 293 and 80 K.}
    \label{Fig1}
\end{figure}

The crystal structure of LuFe$_6$Ge$_6$ is shown in Figs. \ref{Fig1}(a,b). It crystallizes in the HfFe$_6$Ge$_6$-type structure (space group \textit{P6/mmm}, No. 191), featuring a characteristic LEGO-like stacking of Fe$_3$Ge$2$ kagome and honeycomb layers separated by Lu atoms \cite{Kang2020-gv,PhysRevB.111.125163}. X-ray diffraction yields lattice parameters \textit{a}=\textit{b}=5.09(4) Å and \textit{c}=8.07(6) Å at 293 K. Scanning transmission electron microscopy (STEM) images (Figs. \ref{Fig1}(c,d)), acquired in high-angle annular dark-field (HAADF) mode providing atomic-number (Z) contrast, resolve the kagome atomic structure and are in good agreement with the crystallographic structure. The magnetization exhibits an antiferromagnetic transition near room temperature, at T$_\mathrm{N}$=285 K, while the electrical resistivity shows no additional anomaly indicative of a phase transition (Suppl. Inf. Fig. S1) \cite{Teng2022-pf, PhysRevResearch.6.033222}. Furthermore, diffuse-scattering measurements (Figs. \ref{Fig1}(e,f)), performed at 80 and 293 K, reveal only the fundamental Bragg reflections, with no additional satellite reflections detected at either temperature. This excludes a long-range periodic lattice modulation detectable within the sensitivity of the present measurements at these two temperatures. The absence of additional satellite reflections is consistent with the reported harmonic phonon stability of LuFe$_6$Ge$_6$ \cite{feng2024cataloguephononinstabilitiessymmetry} and with previous studies of other {AFe$_6$Ge$_6$} family members \cite{ncm1-rfsn,PhysRevLett.134.186501,Mantravadi2024-ch}. Together, these results provide no evidence for a long-range CDW modulation in LuFe$_6$Ge$_6$ under the investigated conditions, providing the basis for the subsequent electronic structure investigation.

%are $a$ = $b$ = 5.078 Å and $c$ = 8.072 Å at 80 K

\begin{figure}
    \centering
    \includegraphics{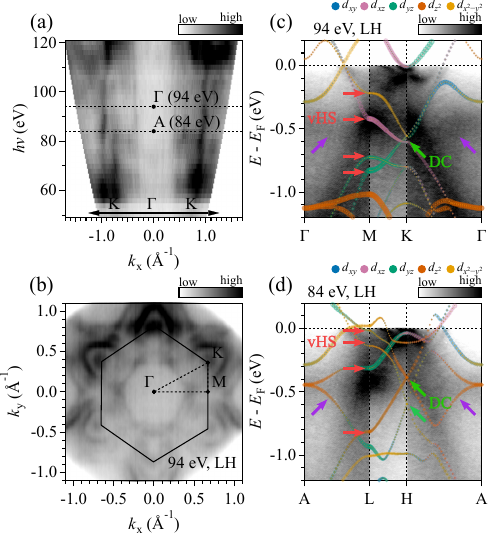}
    \caption{Electronic structure of {LuFe$_6$Ge$_6$} measured by ARPES. (a) Photon-energy-dependent measurement results. Black dashed lines at 94 and 84 eV correspond to the $\Gamma$ and A planes, respectively. (b) Fermi surfaces at the $\Gamma$ plane. Solid lines and dashed lines are the first BZ boundary and high-symmetry directions. (c,d) Band dispersions along (c) $\Gamma$-M-K-$\Gamma$ and (d) A-L-H-A measured with linear horizontal (LH) polarization, respectively. Overlaid markers correspond to the bulk DFT-calculated band structures, with different colors representing different Fe $d$ orbitals.}
    \label{Fig2}
\end{figure}

%The calculated density of states (DOS) peaks at binding energies associated with the flat bands and vHS, Fig. \ref{Fig2}(a). 

The experimental band structure measured by ARPES at 10 K is summarized in Figs. \ref{Fig2} and S3. Details of the ARPES experimental setup and beamline geometry are provided in the Supplementary Information. Photon-energy-dependent scans reveal weak $k_z$ dispersion for states near $E_F$, indicating a predominantly two-dimensional electronic character near the Fermi level, Fig. \ref{Fig2}(a). The Fermi surface at the $\Gamma$ plane, Fig. \ref{Fig2}(b) exhibits a hexagonal topology, with triangular pockets at the K point and a large hexagonal pocket surrounding the $\Gamma$ point \cite{Subires2025-ct,Teng2023-og}. High-symmetry dispersions at the $\Gamma$ and {A} planes are presented in Figs. \ref{Fig2}(c,d) respectively, overlaid with bulk antiferromagnetic DFT calculations. Overall, the measured band structures are in good agreement with DFT, confirming the kagome-derived electronic scaffold. Several vHs are identified at the {M} ({L}) point in both ARPES and DFT (red arrows). In the $\Gamma$ plane, the experimentally observed vHs appear at approximately 0.2 and 0.45 eV below \textit{E}$_F$, with predominantly $d_{x^2-y^2}$ and $d_{xz}$ orbital characters (in global Cartesian frame), respectively, and two additional vHs with $d_{yz}$ character are observed at 0.7 and 0.8 eV, Fig. \ref{Fig2}(c). On the other hand, vHs in the A plane are located approximately at $E_F$ and 0.3 eV below $E_F$ with $d_{yz}$ character, and two additional vHs with $d_{z^2}$ characters appear at 0.15 eV and 0.85 eV below $E_F$, Fig. \ref{Fig2}(d). Similarly, the DCs (green arrows) also exhibit a pronounced $k_z$ dependence. In the {A} plane, ARPES reveals a Dirac point around 0.4 eV below $E_F$, consistent with the DFT predicted $d_{z^2}$ and $d_{x^2-y^2}$ orbital characters, Fig. \ref{Fig2}(d), while it is located 0.6 eV below $E_F$ in the $\Gamma$ plane, Fig. \ref{Fig2}(c). In addition to these bulk-derived bands, flat-like features appear near 0.5 eV (purple arrows), and are attributed to a Ge-derived surface state. This assignment is consistent with the Ge core-level spectrum shown in Fig. S2, DFT-calculated Ge surface bands in Fig. S7, and the Ge surface bands observed in {FeGe} \cite{Subires2025-ct}. The kagome-derived FBs, in contrast, are predicted at 1.5 eV below and 0.5 eV above $E_F$, outside the energy window of the current ARPES measurements (see Fig. S6). Consistent with the absence of CDW order established in Fig. \ref{Fig1}, no band folding or reconstruction is observed. Compared with the bands at higher binding energies, the near-$E_F$ bands exhibit considerably sharper spectral features, as also seen in the momentum distribution curves (MDCs) (Fig. S4). The marked contrast between the sharp near-$E_F$ bands and the substantially broader features at higher binding energies may suggest a change in the nature of the spectral response beyond simple energy-dependent broadening, although its origin cannot be determined from the MDC linewidths alone.

\begin{figure}
    \centering
    \includegraphics{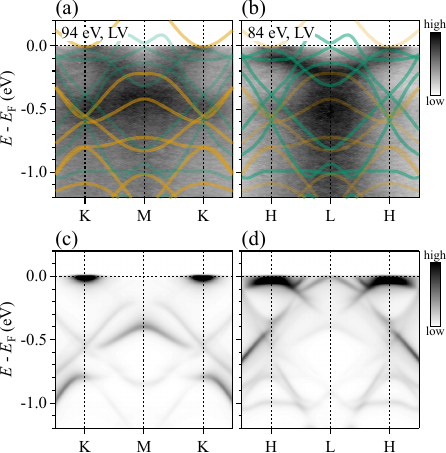}
    \caption{Waterfall feature observed in ARPES. (a,b) ARPES intensity maps measured with LV polarization along {K}-{M}-{K} and {H}-{L}-{H}, respectively. Yellow and green curves overlay the bulk DFT-calculated dispersions along {K}-{M}-{K} ($k_z$ = 0) and {H}-{L}-{H} ($k_z$ = $\pi/c$), respectively. Bands smeared from the other $k_z$ plane due to the $k_z$ broadening effect are shown with increased transparency. (c,d) Calculated ARPES spectral function along {K}-{M}-{K} and {H}-{L}-{H} for {LuFe$_6$Ge$_6$} with linear vertical polarized light.}
    \label{Fig3}
\end{figure}

Parsing the ARPES spectra along the {K}-{M}-{K} ($k_z$ = 0) and {H}-{L}-{H} ($k_z$ = $\pi/c$) directions, we reveal a polarization dependent vertical spectral feature near the {M} and {L} points (Figs. \ref{Fig3}(a-b)). The effect extends beyond 1.2 eV binding energy, forming a characteristic ``waterfall"-like dispersion. Notably, the onset of the vertical spectral feature emerges at approximately 0.3 eV below $E_F$ ({H}-{L}-{H} path) and coincides with the energy of one of the vHs identified in Fig. \ref{Fig2}, suggesting a possible connection between the anomalous spectral weight and the saddle-point band. The waterfall feature has no clear counterpart in the DFT calculations (Fig. S7-S9), while the parent band from which the additional spectral weight emerges is predominantly derived from the Fe $d_{yz}$ orbital, as inferred from its polarization dependence and comparison with the calculations (Figs. \ref{Fig2}(d) and \ref{Fig3}(b)). We must stress that ARPES matrix-element simulations reproduce the polarization-dependence of the underlying bands, but do not generate additional vertical spectral weight at the {L} point, Figs. \ref{Fig3}(c,d).

\begin{figure}
    \centering
    \includegraphics{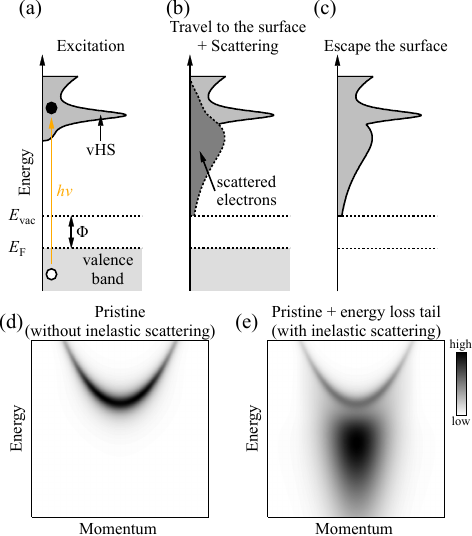}
    \caption{Schematic illustration of inelastic scattering in the photoemission process. (a-c) Energy and photoemission intensity diagrams illustrating the ARPES photoemission process and subsequent inelastic scattering. $E_{\text{vac}}$ and $E_F$ denote the vacuum and Fermi levels, respectively. $h\nu$ is the incident photon energy, and $\Phi$ is the work function. White and black circles represent holes and photoexcited electrons, respectively. (d,e) Schematic ARPES intensity maps showing (d) a pristine parabolic band with enhanced spectral weight at the band bottom due to the vHS and (e) additional spectral weight extending toward higher binding energy due to inelastic energy loss of the photoelectrons. See the main text for details of the model and parameters used in the calculations.}
    \label{Fig4}
\end{figure}

Given that waterfall-like features have frequently been associated with coupling to bosonic excitations and strong electronic correlations, we examined whether such interactions could account for the observed spectral anomaly. The calculated phonon spectrum of LuFe$_6$Ge$_6$ extends to approximately 60 meV \cite{feng2024cataloguephononinstabilitiessymmetry}, substantially below the energy range over which the vertical spectral weight is observed. Although LuFe$_6$Ge$_6$ orders antiferromagnetically at high temperature, its spin-excitation spectrum has not, to our knowledge, been experimentally determined. We therefore cannot establish a corresponding magnon energy scale. Nevertheless, we observe no resolved kink in the measured band dispersions that would provide direct evidence for coupling to a bosonic mode, although this does not exclude such coupling. To assess the low-energy electronic renormalization, we performed DMFT calculations including Hund's coupling and evaluated the quasiparticle weights $Z$ of the Fe 3$d$ orbitals, as summarized in the Suppl. Inf. Table S1. The calculated quasiparticle weights range from $Z=0.759$ to $0.827$, indicating modest low-energy quasiparticle renormalization within the present DMFT calculation. While these quasiparticle weights characterize only the low-energy behavior near $E_F$, the experimentally observed dispersive bands remain in overall good agreement with conventional DFT calculations also at higher binding energies, including the energy range over which the waterfall-like spectral weight develops. Thus, the pronounced vertical spectral weight emerges without a corresponding strong renormalization of the underlying band dispersion, suggesting that the waterfall-like feature cannot be understood solely in terms of strong correlation-induced renormalization.

%Given that waterfall-like features have frequently been associated with the coupling to bosonic excitations and strong electronic correlations, we examined whether such interactions could account for the observed spectral anomaly. First, the phonon spectrum of LuFe$_6$Ge$_6$ extends to $\sim$60 meV \cite{feng2024cataloguephononinstabilitiessymmetry}, while the spin excitation spectrum is limited to 10–50 meV for Fe–Fe exchange and itinerant spin fluctuations \cite{Chen_2024}, rendering the coupling to bosonic modes rather unlikely. To quantify the correlation strength, we performed DMFT calculations including Hund's coupling and evaluated the quasiparticle weights $Z$ of the Fe 3$d$ orbitals, as summarized in the Suppl. Inf. Table S1. The calculated quasiparticle weights are approximately $Z\sim0.8$ for all five Fe 3$d$ orbitals, indicating only moderate quasiparticle renormalization, and placing LuFe$_6$Ge$_6$ in a relatively weakly correlated regime. This is consistent with the overall good agreement between the measured dispersions and conventional DFT calculations, as well as with the relatively weak electronic correlations reported for the kagome metals \cite{Subires2025-ct}. 

These considerations motivate us to explore inelastic scattering of the outgoing photoelectrons as a possible source of the broad additional ARPES intensity. Such extrinsic losses can redistribute primary photoemission intensity toward higher binding energies and thereby contribute additional spectral weight to the measured spectrum \cite{STougaard_PhysRevB.25.4452, JSZhou_pnas.2012625117}. In the present case, the high DOS associated with the vHs provides a large primary photoemission intensity from which such secondary intensity may originate. We emphasize, however, that the high DOS alone does not imply an enhanced inelastic scattering probability, which depends on the momentum- and energy-dependent loss response of the material. Fig. \ref{Fig4} schematically illustrates this process. Following the initial photoexcitation (Fig. \ref{Fig4}(a)), the large primary intensity associated with the vHs provides the initial spectral weight. As the photoelectrons travel toward the surface, a fraction may undergo inelastic scattering and lose kinetic energy and momentum (Fig. \ref{Fig4}(b)), redistributing part of this primary intensity toward higher binding energies, ($E_{\mathrm{ph}},k_{\mathrm{ph}}\rightarrow E_{\mathrm{ph}}-\omega,k_{\mathrm{ph}}-q$) (Fig. \ref{Fig4}(c)). The corresponding effect on the ARPES spectrum is phenomenologically modeled in Figs. \ref{Fig4}(d,e), following \cite{STougaard_PhysRevB.25.4452}:

\begin{equation}
    I_{sec}(E_B) = A\int^{E_B}_0 I_0(E_{B}')K(E_B-E_B')dE_B'
\end{equation}

where $A$ is a dimensionless scaling factor, $E_B$ and $E_B'$ are binding energies, and $I_0$ is the original ARPES intensity. The normalized Gamma-shaped energy-loss kernel is defined as,

\begin{equation}
    K(\Delta E)=\frac{\Delta E}{( E_0)^2}e^{-\Delta E/E_0}
\end{equation}

where $\Delta E=E_B-E_B'$ is the energy loss and $E_0$ sets the characteristic energy-loss scale. The convolution is applied independently at each measured momentum and therefore considers only energy redistribution. We do not explicitly model momentum transfer, as this would require introducing an additional momentum-dependent scattering kernel that is not constrained by the present data. For the illustrative calculations shown in Figs. \ref{Fig4}(d,e), we consider a parabolic dispersion with its band minimum at 0.3 eV below \(E_F\). To mimic the DOS enhancement associated with the vHs, the corresponding spectral function is multiplied by a Lorentzian along the energy axis centered at the vHs energy. We then use \(E_0 = 0.25\) eV and \(A = 1.25\) for the energy-loss contribution. The resulting spectra show how spectral weight originating from the vHs at the bottom of a parabolic band can extend toward higher binding energies, producing a vertical distribution of additional spectral weight. In this scenario, the strong polarization dependence originates from the matrix-element enhancement of the underlying $d_{yz}$ state, which further increases the initial spectral weight available for the inelastic contribution. More broadly, these results suggest that pronounced vertical features in ARPES may contain contributions from extrinsic photoemission processes in addition to intrinsic spectral effects, which should therefore be considered when interpreting such anomalous spectral features.

Our results provide an example of pronounced incoherent spectral weight in a system that is only weakly renormalized within the present DMFT calculation. In {LuFe$_6$Ge$_6$}, the large phase space for electron-hole excitations associated with the enhanced DOS near the vHs may facilitate energy loss of outgoing photoelectrons through excitations within the solid, producing an inelastic contribution to the photoemission spectrum. Furthermore, we observe no resolved `\textit{kink}' in the measured band dispersions that would provide direct evidence for coupling to a bosonic mode. Together, these observations motivate photoelectron energy-loss processes as a possible contribution to the observed vertical spectral weight.

%Our results represent a textbook example of the emergence of incoherent spectral weight in non-correlated materials, that bring important conclusions for the interpretation of vertical spectral features. Keeping aside the controversial waterfalls in superconducting cuprates and pnictides \cite{DSInosov_PhysRevLett.99.237002}, the large phase space for electron–hole excitations close to \textit{E}$_F$ in non-correlated materials, such as kagome metals, allows for energy transfer to excitations inside the solid, producing an inelastic tail in the photoemission spectrum and incoherent quasiparticle peaks. This interpretation should be particularly appealing in the absence of `\textit{kinks}' in the band dispersion that rules out the coupling of electrons to bosonic modes, even though the Hund's coupling is active.

In summary, in contrast to the parent compound FeGe, {LuFe$_6$Ge$_6$} shows no evidence of CDW formation, while its electronic structure agrees with DFT and DMFT. Nevertheless, polarization-dependent ARPES reveals a pronounced vertical spectral feature at the {L} point that onsets near the high DOS associated with the vHs and extends over a broad energy range toward higher binding energies, while the present DMFT calculations indicate only weak electronic renormalization. We propose inelastic scattering of outgoing photoelectrons as a possible origin of the additional spectral weight. Our results suggest that photoelectron energy-loss processes should be considered as a possible contribution when interpreting pronounced vertical spectral features in ARPES.

%In summary, in contrast to the parent compound FeGe, {LuFe$_6$Ge$_6$} shows no evidence of CDW formation, while its electronic structure agrees with DFT and DMFT. Nevertheless, polarization-dependent ARPES reveals a pronounced vertical spectral feature at the {L} point that emerges from a high DOS at the vHs and extends over a broad energy range toward higher binding energies, despite {LuFe$_6$Ge$_6$} being a weakly correlated system. We propose inelastic scattering of photoelectrons as a possible origin of the additional spectral weight. Our results demonstrate that pronounced vertical spectral features can emerge even in absence of electronic correlations, coupling to bosonic modes or Hund´s coupling and highlight the importance of inelastic scattering photoemission processes when interpreting anomalous spectral features in ARPES.

\section*{Acknowledgments}
We thank A. Santander-Syro, Yeongkwan Kim, Ming Yi and Tonica Valla for fruitful discussions. S.B-C. acknowledges financial support from the MINECO of Spain through the projects PID2021-122609NB-C21 and PID2024-161503NB-C21. C.-Y. L. and A. Kar thank the European Union Next Generation EU/PRTR-C17.I1, as well as by IKUR Strategy under the collaboration agreement between Ikerbasque Foundation and DIPC on behalf of the Department of Education of the Basque Government. A.Kar thanks the Basque government for financial support through the project PIBA-2023-1-0051. C.Y., A.S., C.S., and C.F. acknowledge financial support by the Deutsche Forschungsgemeinschaft (DFG, German Research Foundation) through the Würzburg-Dresden Cluster of Excellence ctd.qmat – Complexity, Topology and Dynamics in Quantum Matter (EXC 2147, project-id 390858490). This work was carried out with the support of Diamond Light Source, instrument I05-ARPES (proposal SI36505). 
J.D. acknowledges the computational resources provided by CSC -- IT Center for Science, Finland, and by the Aalto Science-IT project.
J.D. and P.T. were supported by the Finnish Centre of Excellence in Quantum Materials (QMAT) and a collaboration between The Kavli Foundation, Klaus Tschira Stiftung, Kevin Wells, and by the Jane and Aatos Erkko Foundation, the Keele Foundation and the Magnus Ehrnrooth Foundation, as part of the SuperC collaboration.
H.L. and Y.J. were supported by the European Research Council (ERC) under the European Union’s Horizon 2020 research and innovation program (Grant Agreement No. 101020833), as well as by the IKUR Strategy under the collaboration agreement between Ikerbasque Foundation and DIPC on behalf of the Department of Education of the Basque Government. 
B.A.B. was supported by the Gordon and Betty Moore Foundation (grant GBMF8685) towards the Princeton theory program, the Gordon and Betty Moore Foundation’s EPiQS Initiative (grant GBMF11070), the Global Collaborative Network Grant at Princeton University, the Simons Investigator Grant No. 404513, the NSF MRSEC (grant DMR-2011750), the Simons Collaboration on New Frontiers in Superconductivity (grant SFI-MPS-NFS-00006741-01), the Princeton Catalysis Initiative (PCI), the Schmidt Transformative Technology Fund at Princeton University and the NSF through the AI Research Institutes (program award DMR-2433348).

\bibliography{references}

\end{document}

\title{Supplementary Material of 'Incoherent Spectral Weight Emerging from a Van Hove Singularity in a Kagome Metal'}

\author{C. Lim}
\email{chanyoung.lim@dipc.org}
\affiliation{Donostia International Physics Center (DIPC), San Sebastián 20018, Spain}

\author{J. Deng}
\affiliation{Department of Applied Physics, Aalto University School of Science, FI-00076 Aalto, Finland}

\author{A. Korshunov}
\affiliation{Donostia International Physics Center (DIPC), San Sebastián 20018, Spain}

\author{A. Kar}
\affiliation{Donostia International Physics Center (DIPC), San Sebastián 20018, Spain}

\author{D. Subires}
\affiliation{Donostia International Physics Center (DIPC), San Sebastián 20018, Spain}
\affiliation{University of the Basque Country (UPV/EHU), Bilbao 48080, Spain}

\author{H. Li}
\affiliation{Donostia International Physics Center (DIPC), San Sebastián 20018, Spain}

\author{Y. Jiang}
\affiliation{Donostia International Physics Center (DIPC), San Sebastián 20018, Spain}

\author{H. Hu}
\affiliation{Department of Physics, University of Science and Technology of China, Hefei, Anhui 230026, China}

\author{E. Modin}
\affiliation{CIC nanoGUNE, BRTA, San Sebastián 20018, Spain}

\author{P. T\"{o}rm\"{a}}
\affiliation{Department of Applied Physics, Aalto University School of Science, FI-00076 Aalto, Finland}

\author{A. Kumar-Sharma}
\affiliation{Max Planck Institute for Chemical Physics of Solids, Dresden 01187, Germany}

\author{C. Shekhar}
\affiliation{Max Planck Institute for Chemical Physics of Solids, Dresden 01187, Germany}

\author{A. Louat}
\affiliation{Diamond Light Source Ltd, Harwell Science and Innovation Campus, Didcot OX11 0DE, UK}

\author{T. K. Kim}
\affiliation{Diamond Light Source Ltd, Harwell Science and Innovation Campus, Didcot OX11 0DE, UK}

\author{C. Felser}
\affiliation{Max Planck Institute for Chemical Physics of Solids, Dresden 01187, Germany}

\author{B. Andrei Bernevig}
\affiliation{Donostia International Physics Center (DIPC), San Sebastián 20018, Spain}
\affiliation{Department of Physics, Princeton University, Princeton, New Jersey 08544, USA}
\affiliation{IKERBASQUE, Basque Foundation for Science, Bilbao 48013, Spain}

\author{S. Blanco-Canosa}
\email{sblanco@dipc.org}
\affiliation{Donostia International Physics Center (DIPC), San Sebastián 20018, Spain}
\affiliation{IKERBASQUE, Basque Foundation for Science, Bilbao 48013, Spain}

\date{\today}

\maketitle

\section{Sample characterization}

\begin{figure*}
    \centering
    \includegraphics{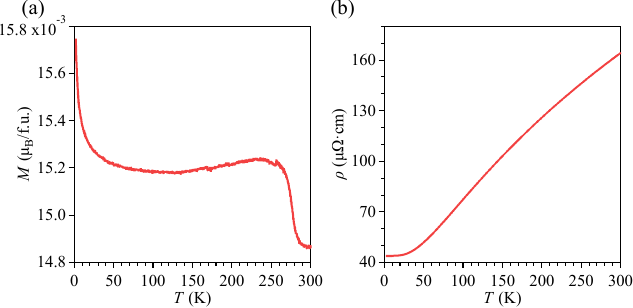}
    \caption{(a) Magnetization and (b) Temperature dependence of the resistivity of LuFe$_6$Ge$_6$. %\purple{Plots need to be updated.}
    }
    \label{Fig_S1}
\end{figure*}

Single crystals of {LuFe$_6$Ge$_6$} were grown by the flux method using high-purity elemental starting materials and characterized by magnetization and resistivity, see Fig. S\ref{Fig_S1}. The magnetization shows an antiferromagnetic transition at 285 K, while the resistivity displays a metallic behavior. Thin specimens for STEM analysis were prepared using the standard focused ion beam (FIB) lift-out technique in a Helios 450S dual-beam scanning electron microscope (Thermo Fisher Scientific, USA). The final cleaning of the thin lamellae was performed in an AmberX (TESCAN, Czech Republic) using a Xe$^+$ ion beam at an energy of 5 keV. STEM measurements were performed using a Titan 60-300 (Thermo Fisher Scientific, USA) operated in HAADF-STEM mode at an accelerating voltage of 300 kV. Diffuse scattering (DS) maps were collected at the European Synchrotron Radiation Facility on the ID28 beamline, employing an incident photon energy of 17.8 keV and a Dectris PILATUS3 1M area detector. Polarization dependence Angle-resolved photoemission spectroscopy (ARPES) measurements were performed at the Diamond Light Source on the I05 beamline. Single-crystal samples were mounted using conductive silver epoxy, with a ceramic top post attached to enable \textit{in-situ} cleaving. Measurements were conducted at a base pressure of 5 $\times$ 10$^{-11}$ torr and a temperature of 10 K. The total energy resolution was 20 meV or better, and the angular resolution was approximately 0.1$^\circ$. The Scienta R4000 analyzer was operated with the entrance slit oriented vertically. Linear horizontal (LH) and linear vertical (LV) polarized light was used to probe the polarization dependence of the electronic structure.

\section{Angle-resolved (ARPES) and Core Level photoemission spectroscopy (XPS)}
Prior to ARPES measurements, a core-level spectrum of LuFe$_6$Ge$_6$ was measured with 70 eV photons (Fig. S2). The spectrum exhibits well-defined Lu, Fe, and Ge peaks, including two Lu 4$f$ peaks at 7 and 8 eV below the Fermi level ($E_F$), and Ge 3$d$ peaks at 29-30 eV. The latter can be fitted with four components (inset), with a line shape resembling that reported for Ge-terminated FeGe \cite{Teng2023-og}, consistent with a similar surface termination.

\begin{figure*}
    \centering
    \includegraphics{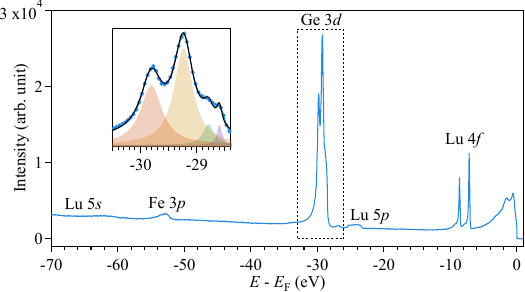}
    \caption{Core level spectrum of {LuFe$_6$Ge$_6$}. The inset shows a magnified view of the region indicated by the dashed box. Blue markers and black solid lines represent the experimental data and the fitted curve, respectively. The four shaded peaks correspond to the individual Lorentzian peaks used in the fitting.}
    \label{Fig_S2}
\end{figure*}

\begin{figure*}
    \centering
    \includegraphics{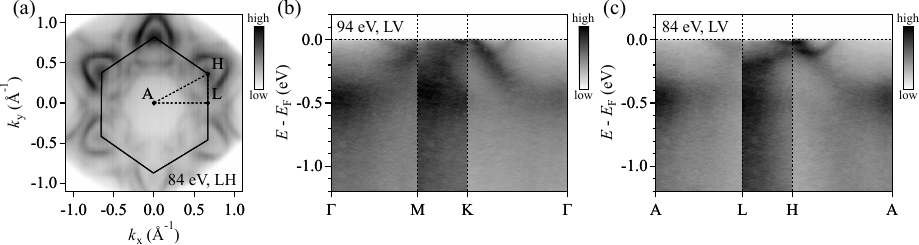}
    \caption{Additional ARPES intensity maps. (a) A-plane Fermi surface measured with 84 eV photons under LH polarization. (b,c) High-symmetry dispersions along the (b) $\Gamma$-M-K-$\Gamma$ and (c) A-L-H-A directions measured under LV polarization, respectively.}
    \label{Fig_S3}
\end{figure*}

Fig. S\ref{Fig_S3} shows additional ARPES results of {LuFe$_6$Ge$_6$}. Fig. S\ref{Fig_S3}(a) presents the Fermi surface at the A plane measured with 84 eV LH photons. The overall topology is nearly identical to that of the $\Gamma$ plane shown in Fig. 2(b) of the main text, indicating weak $k_z$ dependence of the Fermi surface, in contrast to the stronger $k_z$ dispersion observed at higher binding energies. Fig. S\ref{Fig_S3}(b) and S\ref{Fig_S3}(c) show band dispersions measured under LV polarization along the (b) $\Gamma$-M-K-$\Gamma$ and (c) A-L-H-A directions. The waterfall features at the M and L points are visible in these measurements, although they are most clearly observed when the spectra are taken along the K-M-K and H-L-H directions due to matrix-element effects.

\begin{figure*}
    \centering
    \includegraphics{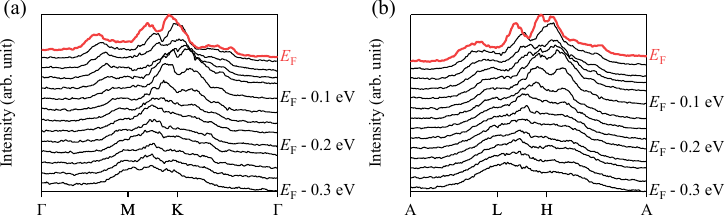}
    \caption{Momentum distribution curves (MDCs) along the (a) $\Gamma$-M-K-$\Gamma$ and (b) A-L-H-A directions over the energy range from 0.3 eV below $E_F$ to $E_F$, measured with LH polarization. The thicker red curves correspond to the MDCs at $E_F$.}
    \label{Fig_S4}
\end{figure*}

Fig. S\ref{Fig_S4} displays momentum distribution curves (MDCs) along the $\Gamma$-M-K-$\Gamma$ and A-L-H-A directions over the binding-energy range from $E_F$ to 0.3 eV below $E_F$. The MDCs at $E_F$, highlighted by the thicker red curves, exhibit sharp and well-defined peaks, providing evidence for good sample quality and indicating that the pronounced broadening observed at higher binding energies is unlikely to originate solely from sample-related disorder. With increasing binding energy, the MDC features become progressively broader, revealing a clear energy-dependent evolution of the spectral linewidth, that may be related to the development of the observed vertical spectral feature at the {L} point.

\section{Density Functional Theory (DFT)}

Density-functional theory (DFT) calculations were performed with the projector augmented-wave (PAW) method~\cite{Blochl1994ProjectorA,Kresse1999FromA} as implemented in the Vienna ab initio simulation package (\texttt{VASP})~\cite{Kresse1996EfficiencyA,Kresse1996EfficientA}, using the Perdew--Burke--Ernzerhof (PBE) generalized gradient approximation~\cite{Perdew1996GeneralizedA} and keeping the Lu $4f$ states in the core.
The plane-wave cutoff was 500\,eV, the Brillouin zone was sampled on a $\Gamma$-centered $12\times12\times6$ mesh~\cite{Monkhorst1976SpecialA}. {LuFe$_6$Ge$_6$} was treated in the HfFe$_6$Ge$_6$-type structure (space group $P6/mmm$, No.~191) in the collinear A-type antiferromagnetic state, without spin--orbit coupling and without a Hubbard $U$, with the lattice parameters relaxed to $a=5.0814$\,\AA{} and $c=8.1181$\,\AA.
A tight-binding Hamiltonian was constructed from the DFT bands with \textsc{Wannier90}~\cite{Mostofi2008Wannier90A,Pizzi2020Wannier90A} through its \texttt{VASP} interface~\cite{Franchini2012MaximallyA}, using Fe-$d$ and Ge-$p$ orbitals defined along the global Cartesian axes, with $\hat{x}\parallel\Gamma$--K and $\hat{z}\parallel c$.
Orbital-resolved bulk bands and the spectral functions of the semi-infinite (001) surface for both terminations of the layer stacking, the Ge hexagonal and the Fe kagome layer, obtained with the iterative Green's-function method~\cite{Sancho1984QuickA,Sancho1985HighlyA}, were computed from this Hamiltonian with \textsc{WannierTools}~\cite{Wu2018WannierToolsA}.
Polarization-dependent ARPES intensities were simulated with \texttt{chinook}~\cite{Day2019ComputationalA}.
The light was taken to be incident in the $yz$ plane at 45$^\circ$ from the surface normal, with LV polarized along the analyser slit ($\hat{x}\parallel\Gamma$--K) and LH along $(\hat{y}+\hat{z})/\sqrt{2}$, reproducing the polarization directions of the experiment.

\begin{figure*}
    \centering
    \includegraphics{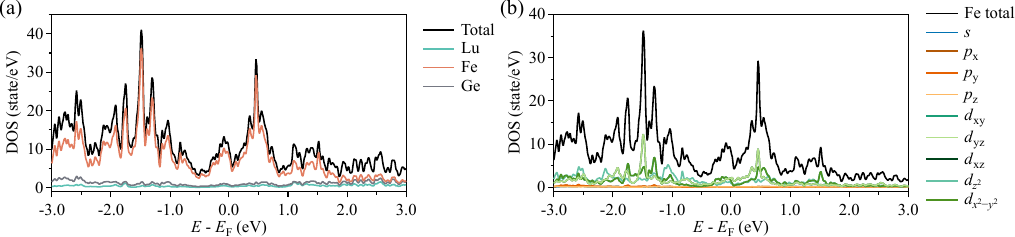}
    \caption{DOS calculated by DFT. (a) TDOS and PDOS for each atom in {LuFe$_6$Ge$_6$}. (b) PDOS for Fe, including contributions from the total Fe states and the Fe $s$, $p$, and $d$ orbitals.}
    \label{Fig_S5}
\end{figure*}

\begin{figure*}
    \centering
    \includegraphics{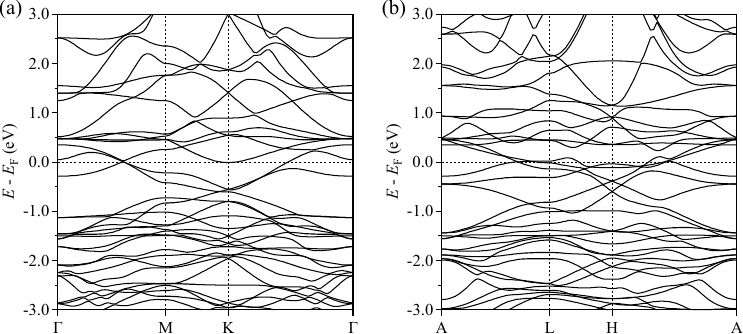}
    \caption{(a,b) DFT-calculated bulk electronic structure of {LuFe$_6$Ge$_6$} over a larger energy window along the (a) $\Gamma$-M-K-$\Gamma$ and (b) A-L-H-A directions, respectively.}
    \label{Fig_S6}
\end{figure*}

Fig. S\ref{Fig_S5} shows the density of states (DOS) of {LuFe$_6$Ge$_6$} calculated using DFT. Fig. S\ref{Fig_S5}(a) presents the total DOS (TDOS) together with the atomic contributions. The Fe contribution dominates near $E_F$. Two pronounced peaks appear at 1.5\,eV below and 0.5\,eV above $E_F$, corresponding to the FBs. The detailed band dispersions associated with these features are shown in Fig. S\ref{Fig_S6}. Fig. S\ref{Fig_S5}(b) shows the projected DOS (PDOS) of the Fe atom, including the contributions from individual Fe orbitals. The DOS peaks corresponding to the FBs exhibit strong $d_{yz}$ orbital character.

\begin{figure*}
    \centering
    \includegraphics{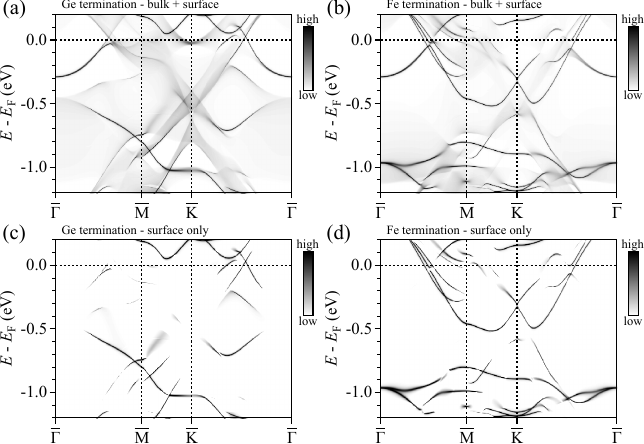}
    \caption{(a,b) DFT-calculated total spectral weight of {LuFe$_6$Ge$_6$} along the $\overline{\Gamma}-\overline{\mathrm{M}}-\overline{\mathrm{K}}-\overline{\Gamma}$ direction for (a) Ge hexagonal termination and (b) Fe kagome termination, respectively. (c,d) Surface-only spectral weight along the $\overline{\Gamma}-\overline{\mathrm{M}}-\overline{\mathrm{K}}-\overline{\Gamma}$ direction for (c) Ge hexagonal termination and (d) Fe kagome termination, respectively.} 
    \label{Fig_S7}
\end{figure*}

The calculated (001) surface spectrum for the Ge-terminated surface of {LuFe$_6$Ge$_6$} is shown in Fig. S\ref{Fig_S7}. Most of the prominent dispersive features are reproduced by the surface calculation, whereas the waterfall-like structure observed experimentally under LV polarization near a photon energy of about 84\,eV, corresponding approximately to the $k_z=\pi$ plane, is not captured within the bare band description alone. This comparison already suggests that the anomaly is not simply an additional surface band.

Near the Fermi level, the spectral weight contains a substantial contribution from bulk-projected states associated with the more out-of-plane Fe $3d$ orbitals, especially $d_{xz}$, $d_{yz}$, and $d_{z^2}$. Because these orbitals hybridize strongly along the crystallographic $c$ direction, they also provide the most natural channel through which finite $k_z$ resolution could smear the measured ARPES intensity. This motivates the explicit $k_z$-broadening check discussed next. Because {LuFe$_6$Ge$_6$} is a three-dimensional material, a natural first question is whether the waterfall-like feature could arise from unresolved $k_z$ integration. %The broadened DFT spectra constructed from the procedure above are shown in Fig. S\ref{Fig_S8}(a-j). The calculated \textit{k}$_z$ broadened spectra show a smooth broadening of the bands, but does not reproduce the incoherent spectral weigth.

\section{Analysis on the vertical feature}
\subsection{\textit{k}$_z$ broadening}

\begin{figure*}
    \centering
    \includegraphics{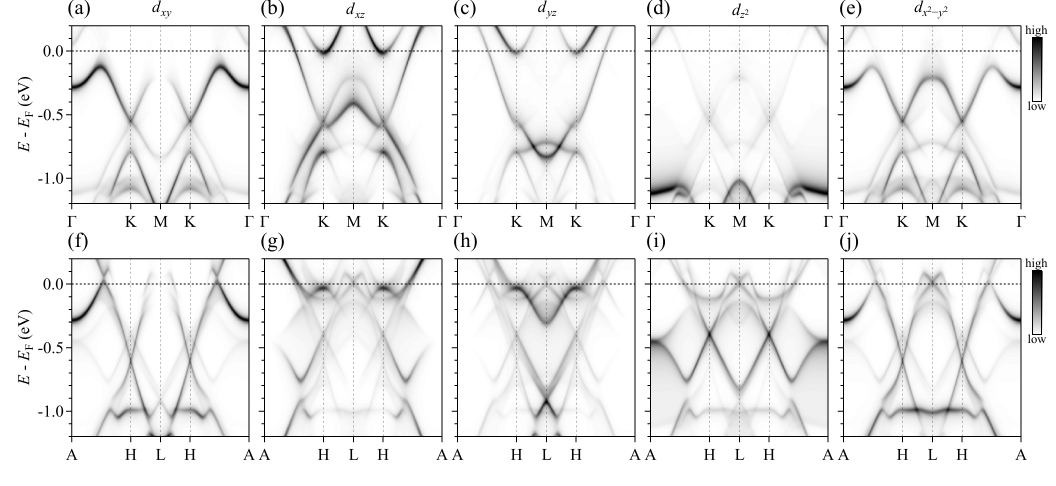}
    \caption{Orbital-resolved $k_z$-broadened bulk spectral function of {LuFe$_6$Ge$_6$}.
    The five columns show the spectral weight projected onto $d_{xy}$, $d_{xz}$, $d_{yz}$, $d_{z^2}$, and $d_{x^2-y^2}$, respectively, as labelled on the panels.
    The top row shows the broadened spectra along $\Gamma$--K--M--K--$\Gamma$ ($k_{z,0}=0$), the bottom row the corresponding spectra along A--H--L--H--A ($k_{z,0}=\pi/c$).
    All ten panels share a single colour scale, normalized to the largest value over the whole set.
    The broadened spectra show that finite $k_z$ resolution produces a visible but comparatively smooth broadening of the underlying bands.
    While this effect can reduce spectral contrast, it does not reproduce the abrupt loss of coherence associated with the observed waterfall-like feature.
    Finite $k_z$ broadening is therefore unlikely to be the primary origin of the anomaly.} 
    \label{Fig_S8}
\end{figure*}

Because ARPES has a finite out-of-plane momentum resolution, the measured intensity in a three-dimensional material reflects a weighted average over a finite range of $k_z$ rather than a perfectly sharp cut at a single $k_z$ plane.
To simulate this effect, the spectral intensity at a target plane $k_{z,0}$ is constructed by combining the spectra computed for a set of discrete sampled $k_z$ slices.
For each slice $k_z$ and band index $n$, the contribution at in-plane momentum $k_{\parallel}$ and energy $\omega$ is weighted both by the orbital-projection factor and by a broadening kernel centered at $k_{z,0}$.
Explicitly,
\begin{equation}
I_g(k_{\parallel},\omega;k_{z,0})
=
\sum_{k_z}\sum_n
W_g(k_{\parallel},n;k_z)\,
K_z(k_z-k_{z,0})\,
K_E\!\left[\omega-\varepsilon_n(k_{\parallel},k_z)\right],
\label{eq:supp:kz-broadened-intensity}
\end{equation}
with
\begin{align}
W_g(k_{\parallel},n;k_z)
&=
\sum_{\alpha \in g}
\left|\langle \alpha \mid \psi_n(k_{\parallel},k_z)\rangle\right|^2,\\
K_z(\Delta k_z)
&=
\frac{\Gamma_z/\pi}{(\Delta k_z)^2+\Gamma_z^2},
\qquad
K_E(\Delta \omega)
=
\frac{\eta/\pi}{(\Delta \omega)^2+\eta^2},
\label{eq:supp:kz-weight-and-kernels}
\end{align}
Here $I_g(k_{\parallel},\omega;k_{z,0})$ is the simulated spectral intensity for orbital group $g$ at in-plane momentum $k_{\parallel}$, energy $\omega$, and target out-of-plane momentum $k_{z,0}$.
The sum over $k_z$ runs over the discrete sampled out-of-plane slices included in the calculation, and the sum over $n$ runs over the bands at each sampled slice.
The quantity $W_g(k_{\parallel},n;k_z)$ is the orbital weight of band $n$ in orbital group $g$ at $(k_{\parallel},k_z)$, and the index $\alpha$ labels the individual orbitals belonging to the chosen group $g$.
Here $\psi_n(k_{\parallel},k_z)$ is the eigenstate of band $n$ at the sampled slice $k_z$ and $\varepsilon_n(k_{\parallel},k_z)$ is the corresponding band energy.
In the kernel definitions, $\Delta k_z \equiv k_z-k_{z,0}$ is the offset of a sampled slice from the target plane, and $\Delta \omega \equiv \omega-\varepsilon_n(k_{\parallel},k_z)$ is the energy offset from the band energy.
$\Gamma_z$ is the half width at half maximum of the $k_z$ kernel and $\eta$ that of the energy kernel.
In the present calculation $\Gamma_z=0.1\times2\pi/c=0.077$\,\AA$^{-1}$ and $\eta=20$\,meV~\cite{Strocov2003IntrinsicA}, evaluated on a 5\,meV energy grid.
The sum over $k_z$ runs over 51 slices spanning $k_z=0$ to $\pi/c$ in steps of $0.01\times2\pi/c$, using $\varepsilon_n(-k_z)=\varepsilon_n(k_z)$.

To test this possibility, we applied the finite-$k_z$ procedure to the orbital-resolved bulk spectra.
This step is particularly relevant for {LuFe$_6$Ge$_6$} because the identified out-of-plane Fe $3d$ orbitals disperse strongly along $k_z$ and should therefore be the most sensitive to unresolved out-of-plane averaging.
The resulting $k_z$-broadened spectra are shown in Fig.~S\ref{Fig_S8}.
While the finite-$k_z$ average visibly smears the band contrast, it does not generate the characteristic abrupt loss of coherent dispersion observed experimentally.
We therefore conclude that $k_z$ broadening alone is insufficient to account for the anomaly.

\begin{figure*}
    \centering
    \includegraphics{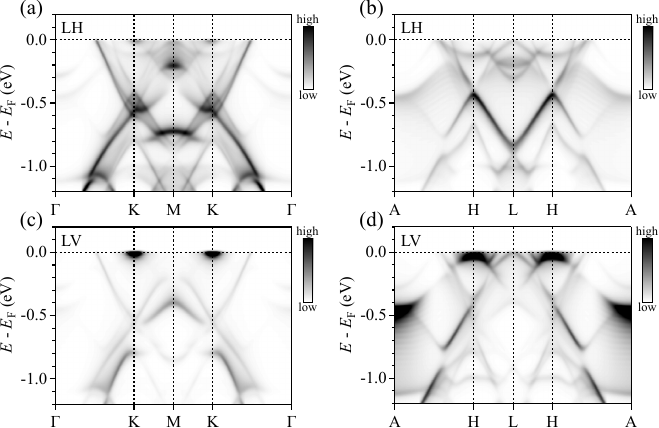}
    \caption{ARPES intensity simulation taking into account the matrix elements. (a,b) LH and (c,d) LV polarization, along (a,c) $\Gamma$--K--M--K--$\Gamma$ at $k_z=0$ and (b,d) A--H--L--H--A at $k_z=\pi/c$.
    The two $k_z$ planes are imposed directly rather than converted from photon energy, so no inner potential enters.}
    \label{Fig_S9}
\end{figure*}

Finally, the ARPES intensity including the matrix-element effect is plotted in Fig.~S\ref{Fig_S9}.
The calculations, performed with the \texttt{chinook} code~\cite{Day2019ComputationalA}, do not reproduce the waterfall feature.  

\subsection{DMFT}

Waterfall-like anomalies are often discussed in strongly correlated systems, especially cuprates, where many-body renormalization strongly reshapes the low-energy spectral function~\cite{Damascelli_2003,Sobota_2021}.
We therefore examined whether the anomaly in {LuFe$_6$Ge$_6$} could be interpreted in the same way.

As a simple measure of correlation strength, we evaluated the DMFT quasiparticle weight for the Fe $3d$ orbitals in the AFM phase.
Within DMFT, the self-energy is taken to be local,
$\Sigma(\mathbf{k},\omega)\rightarrow\Sigma(\omega)$, so the quasiparticle renormalization is determined by the frequency dependence of the local self-energy.
Starting from the retarded Green's function
\begin{equation}
G^{-1}(\mathbf{k},\omega)=\omega+\mu-\varepsilon_{\mathbf{k}}-\Sigma^R(\omega),
\end{equation}
we expand the self-energy near the Fermi level,
\begin{equation}
\Sigma^R(\omega)\approx
\Sigma^R(0)+
\omega\left.\frac{\partial \Sigma^R(\omega)}{\partial \omega}\right|_{\omega=0}.
\end{equation}
Keeping only the linear term, the coefficient of $\omega$ is renormalized by the self-energy slope, which gives the quasiparticle weight
\begin{equation}
Z=
\left[
1-\left.\frac{\partial \Re \Sigma^R(\omega)}{\partial \omega}\right|_{\omega=0}
\right]^{-1}.
\label{eq:dmft_Z_real}
\end{equation}

The same quantity can be obtained on the Matsubara axis.
For a Fermi liquid, the self-energy is analytic near zero frequency, so
\begin{equation}
\Sigma(i\omega_n)\approx
\Sigma(0)+
i\omega_n
\left.\frac{\partial \Sigma(i\omega_n)}{\partial (i\omega_n)}\right|_{\omega_n\to 0},
\end{equation}
which implies
\begin{equation}
Z=
\left[
1-\left.\frac{\partial \Im \Sigma(i\omega_n)}{\partial \omega_n}\right|_{\omega_n\to 0}
\right]^{-1}.
\label{eq:dmft_Z_matsubara}
\end{equation}
At finite temperature the fermionic Matsubara frequencies $\omega_n=(2n+1)\pi T$ are discrete, with $k_\mathrm{B}=1$, and this derivative is estimated from the first two of them,
\begin{equation}
Z\approx
\left[
1-\frac{\Im \Sigma(i\omega_1)-\Im \Sigma(i\omega_0)}{\omega_1-\omega_0}
\right]^{-1}.
\label{eq:dmft_Z_two_point}
\end{equation}
In the present DMFT analysis, we use Eq.~\eqref{eq:dmft_Z_two_point} to estimate the quasiparticle weight of the Fe $3d$ orbitals.
The DMFT quasiparticle weights $Z$ are summarized in Table~S\ref{tab:dmft_Z_orbitals}, with quasiparticle weights of order $Z\sim0.8$ for all Fe $3d$ orbitals.
This level of renormalization is consistent with a weakly correlated multiorbital metal and suggests the strong correlation might not be the primary mechanism for the observed waterfall feature in {LuFe$_6$Ge$_6$}.

Electronic correlations in the Fe $3d$ shell were assessed with one-shot DFT plus dynamical mean-field theory (DFT+DMFT)~\cite{Georges1996DynamicalA,Kotliar2006ElectronicA} using \texttt{solid\_dmft}~\cite{Merkel2022SoliddmftA}, built on the TRIQS library~\cite{Parcollet2015TRIQSA} and TRIQS/DFTTools~\cite{Aichhorn2016TRIQSDFTToolsA}.
The DFT part was a \texttt{VASP} calculation with the same functional, potentials, cutoff and A-type antiferromagnetic order as above, on a $\Gamma$-centered $9\times9\times6$ mesh.
Projected localized orbitals~\cite{Schuler2018ChargeA} were built from the PAW projectors for the Fe-$3d$ orbitals (correlated) and the Ge-$4p$ orbitals (uncorrelated).
The two inequivalent Fe shells are the two kagome layers of the A-type order.
The local interaction was of density--density type with $U=4.2$\,eV and $J_\mathrm{H}=0.88$\,eV as in Ref.~\cite{Miao2023SignatureA}, the double counting was treated in the fully localized limit~\cite{Anisimov1993DensityfunctionalA} using the DFT occupations, and the impurity problem was solved with the continuous-time hybridization-expansion quantum Monte Carlo method~\cite{Werner2006ContinuousTimeA,Gull2011ContinuoustimeA} of TRIQS/CTHYB~\cite{Seth2016TRIQSCTHYBA} at an inverse temperature $\beta=100$\,eV$^{-1}$.
Quasiparticle weights were estimated from Eq.~\eqref{eq:dmft_Z_two_point} without analytic continuation.
At $\beta=100$\,eV$^{-1}$ the temperature is $T=116$\,K, well above the 10\,K of the ARPES measurement.

\begin{table}[t]
    \centering
    \caption{Estimated quasiparticle weights $Z$ for the Fe $3d$ orbitals from the DMFT self-energy, evaluated with the two-point Matsubara slope of Eq.~\eqref{eq:dmft_Z_two_point}.
The two impurity problems are the two kagome layers of the A-type order and are related by interchange of the spin channels.}
    \label{tab:dmft_Z_orbitals}
    \begin{tabular}{c c c c c c c}
        \hline\hline
        Impurity & Spin & $d_{xy}$ & $d_{yz}$ & $d_{z^2}$ & $d_{xz}$ & $d_{x^2-y^2}$ \\
        \hline
        0 & up   & 0.776 & 0.759 & 0.765 & 0.766 & 0.765 \\
        0 & down & 0.803 & 0.817 & 0.813 & 0.805 & 0.827 \\
        1 & up   & 0.803 & 0.817 & 0.813 & 0.805 & 0.827 \\
        1 & down & 0.776 & 0.759 & 0.765 & 0.766 & 0.765 \\
        \hline\hline
    \end{tabular}
\end{table}

% \begin{figure*}
%     \centering
%     \includegraphics{figure_s10.pdf}
%     \caption{The first row shows different forms of self-energy, the second and third rows are the $A_r(k,\omega)$ function the spectral function $A(k,\omega)$ calculated from the self-energy in the first row in the same column. The real part of self-energy is obtained from the imaginary part by Kramers-Kronig relation, and then subtract a constant to make the real part be zero when $\omega=0$. The bare dispersion has parameters $t=0.5,\ t_0=0.3$ with chemical potential $\mu=0$. The bare system with zero self-energy is in (a), and the MFL and FL are in (b) and (c) respectively with $E_{sat}=0.65,\ \omega_1=0.5,\ \omega_2=1$. (d,e) represent other forms of self-energy. Waterfall features can be observed in the spectral function in the third row.}
%     \label{Fig_S10}
% \end{figure*}

\subsection{Self-energy mechanism for a waterfall-like spectral feature}

This subsection considers a generic intrinsic self-energy mechanism as one possible phenomenological route to a waterfall-like spectral feature. Using the notation introduced above and neglecting photoemission matrix-element variations, the ARPES intensity along a one-dimensional momentum cut is proportional to the spectral function
\begin{align}
A(k,\omega)&=-\frac{1}{\pi}\Im G(k,\omega),\\
G^{-1}(k,\omega)&=\omega+\mu-\varepsilon_{k}-\Sigma^R(\omega).
\end{align}
For a momentum-independent self-energy, this gives
\begin{equation}
A(k,\omega)=\frac{1}{\pi}
\frac{-\Im \Sigma^R(\omega)}
{[\omega+\mu-\varepsilon_{k}-\Re \Sigma^R(\omega)]^2
+[\Im \Sigma^R(\omega)]^2}.
\label{eq:supp:spectral-function}
\end{equation}
For a one-dimensional momentum cut at fixed $\omega$, let $k^*(\omega)$ denote the position of the momentum distribution curve (MDC) maximum. Away from a bare-band extremum, it satisfies
\begin{equation}
\omega+\mu-\varepsilon_{k^*}-\Re \Sigma^R(\omega)=0.
\label{eq:supp:kstar}
\end{equation}
Differentiating Eq.~\eqref{eq:supp:kstar} with respect to $\omega$ yields
\begin{equation}
\frac{d k^*}{d\omega}=
\frac{1-\partial\Re \Sigma^R(\omega)/\partial\omega}
{d\varepsilon_{k^*}/d k^*}.
\label{eq:supp:dkstar}
\end{equation}
The waterfall-like spectral feature is characterized by an MDC-peak locus that changes little over an extended energy interval, $d k^*/d\omega\approx0$. Within the present local-self-energy model, the corresponding condition is therefore
\begin{equation}
\frac{\partial\Re \Sigma^R(\omega)}{\partial\omega}\approx1.
\label{eq:supp:waterfall-condition}
\end{equation}
Equation~\eqref{eq:supp:waterfall-condition} describes the nearly vertical locus of MDC maxima in the intensity map. The relevant quantity is the finite-frequency slope of $\Re \Sigma^R$. The real and imaginary parts of a causal retarded self-energy are related by the Kramers--Kronig relation
\begin{equation}
\Re \Sigma^R(\omega)=\frac{1}{\pi}\,\mathcal{P}
\int_{-\infty}^{\infty}
\frac{\Im \Sigma^R(\omega')}{\omega'-\omega}\,d\omega'.
\label{eq:supp:kk}
\end{equation}
We can examine whether the waterfall feature can emerge for some typical forms of self-energy like the Fermi liquid (FL) or marginal Fermi liquid (MFL), with the imaginary parts having the forms
\begin{align}
-\Im \Sigma_{\mathrm{FL}}^R(\omega)&\propto \omega^2+T^2,\\
-\Im \Sigma_{\mathrm{MFL}}^R(\omega)&\propto \max(|\omega|,T).
\label{eq:supp:fl-mfl}
\end{align}
These low-energy expressions cannot increase indefinitely. To represent a finite interval of enhanced inelastic scattering on the occupied side, we use the simple ansatz
\begin{equation}
\Im \Sigma^R(\omega)=
\begin{cases}
-E_{\mathrm{sat}}|\omega|^\alpha/\omega_1^\alpha,
&-\omega_1<\omega\leq0,\\
-E_{\mathrm{sat}},
&-\omega_2\leq\omega\leq-\omega_1,\\
-E_{\mathrm{sat}}\omega_2^2/\omega^2,
&\omega<-\omega_2,
\end{cases}
\label{eq:supp:self-energy-ansatz}
\end{equation}
where $\alpha=2$ and $1$ reproduce the FL and MFL low-energy power laws, respectively. The real part is obtained from Eq.~\eqref{eq:supp:kk}.

\begin{figure*}[t]
    \centering
    \includegraphics{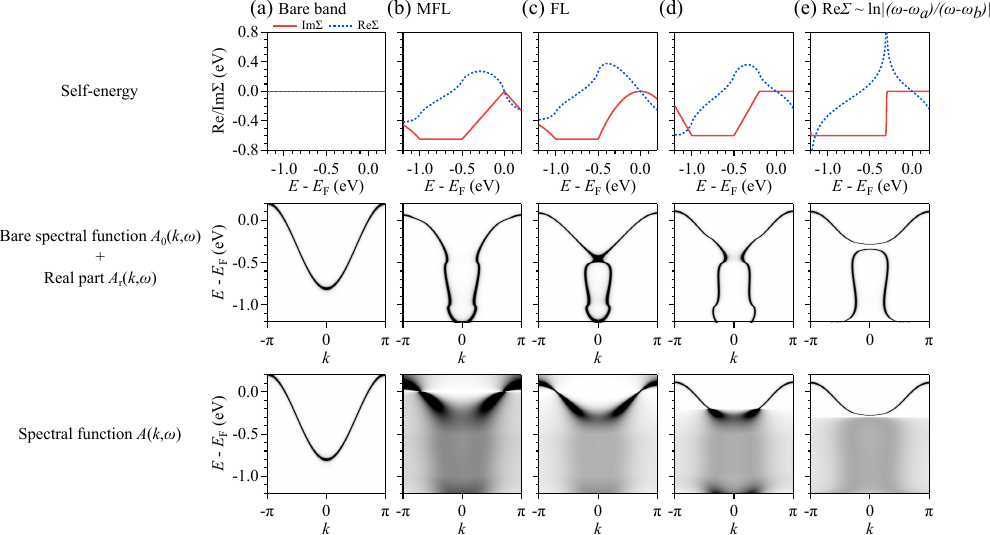}
    \caption{Model retarded self-energies and the resulting waterfall-like spectral functions. The top row shows $\Im \Sigma^R(\omega)$ and the Kramers--Kronig-related $\Re \Sigma^R(\omega)$, shifted so that $\Re \Sigma^R(0)=0$. The middle row shows the auxiliary function $A_r(k,\omega)$, obtained by setting $-\Im \Sigma^R(\omega)=\delta>0$ in Eq.~\eqref{eq:supp:spectral-function} while retaining $\Re \Sigma^R(\omega)$; it therefore highlights the MDC-peak locus $k^*(\omega)$. The bottom row shows the full spectral function $A(k,\omega)$. The columns correspond to (a) the bare band, (b) an MFL self-energy, (c) an FL self-energy. Panels (d) and (e) illustrate the progression from the FL/MFL-like self-energy toward the analytically transparent limit in (e), isolating how enhanced scattering over a finite frequency interval produces a positive-slope region in $\Re \Sigma^R$. The parameters are $t=0.5$~eV, $t_0=0.3$~eV, $\mu=0$, $E_{\mathrm{sat}}=0.65$~eV, $\delta=0.001$ eV, $\omega_1=0.5$~eV, and $\omega_2=1$~eV. A nearly vertical MDC-peak locus develops where $\partial\Re \Sigma^R/\partial\omega$ approaches unity.}
    \label{Fig_S10}
\end{figure*}

We examine whether the waterfall feature appears for the FL and MFL self-energies using a simple toy model with bare dispersion along the momentum cut being $\varepsilon_k=-t\cos k-t_0$. The spectral function obtained by different ansatzs of self-energy are shown in Fig.~\ref{Fig_S10}. The spectral function $A(k,\omega)$ is calculated by Eq.\eqref{eq:supp:spectral-function}, and we also calculate another spectral function $A_r(k,\omega)$ obtained by setting the imaginary part of self-energy in Eq.\eqref{eq:supp:spectral-function} to be a small number $\delta$ while keeping the real part intact. The function $A_r(k,\omega)$ for each $\omega$ is peaked when $k$ is close to $k^*$ defined by Eq.\eqref{eq:supp:kstar}, therefore it tracks the location of $k^*$ as a function of $\omega$. When $k^*$ does not change with $\omega$, it indicates the emergence of waterfall, which manifests as a vertical segment in the $A_r(k,\omega)$. Fig.~\ref{Fig_S10}(a) shows the bare dispersion without self-energy, and in Fig.~\ref{Fig_S10}(b,c) the waterfall feature can emerge after adding the MFL or FL self-energies. The waterfall occurs when the real part of self energy has positive derivative versus $\omega$ close to 1. To find how this positive derivative appears, we further deform the imaginary part into (d) and (e), where in (e) the imaginary part is a finite constant $\Im \Sigma^R(\omega)=-\Gamma_{\mathrm{im}}$ only in some window $\omega_a<\omega<\omega_b$. In this case the waterfall and the linear positive slope in the real part of self energy still persist, and the real part can be obtained analytically as  
\begin{align}
\Re \Sigma^R(\omega)
&=\frac{1}{\pi}\,\mathcal{P}\!\int_{\omega_a}^{\omega_b}
\frac{-\Gamma_{\mathrm{im}}}{\omega'-\omega}\,d\omega'+C\\
&=\frac{\Gamma_{\mathrm{im}}}{\pi}
\ln\left|\frac{\omega-\omega_a}{\omega-\omega_b}\right|+C,
\label{eq:supp:rectangular-self-energy}
\end{align}
where $C$ is a frequency-independent constant. Inside the region $\omega_a<\omega<\omega_b$, $\Re \Sigma^R$ contains a positive-slope region that can satisfy the condition in Eq.~\eqref{eq:supp:waterfall-condition} for the waterfall feature to appear.

The same behavior can be parameterized by a sum of broadened poles,
\begin{equation}
\Sigma^R(\omega)=\Sigma_\infty^R+
\sum_j f_j^R(\omega),\qquad
f_j^R(\omega)=\frac{a_j}{\omega-E_j+i\Gamma_j},
\qquad a_j>0,\quad\Gamma_j>0.
\label{eq:supp:pole-expansion}
\end{equation}
For one pole,
\begin{align}
\Re f_j^R(\omega)
&=\frac{a_j(\omega-E_j)}{(\omega-E_j)^2+\Gamma_j^2},\\
\Im f_j^R(\omega)
&=-\frac{a_j\Gamma_j}{(\omega-E_j)^2+\Gamma_j^2},\\
\frac{\partial\Re f_j^R(\omega)}{\partial\omega}
&=\frac{a_j[\Gamma_j^2-(\omega-E_j)^2]}
{[(\omega-E_j)^2+\Gamma_j^2]^2}.
\label{eq:supp:pole-components}
\end{align}
The derivative is positive for $|\omega-E_j|<\Gamma_j$ and reaches its maximum value $a_j/\Gamma_j^2$ at $\omega=E_j$. A single broadened pole can therefore meet the waterfall condition only when $a_j/\Gamma_j^2\gtrsim1$; for several poles, their contributions to $\partial\Re \Sigma^R/\partial\omega$ add. This pole expansion is a convenient phenomenological representation of a structured scattering rate.

Taken together, the panels in Fig.~S\ref{Fig_S10} illustrate a common mechanism for the waterfall-like spectral feature. In the sharp limit of Fig.~S\ref{Fig_S10}(e), a box-like enhancement of $-\Im \Sigma^R$ between $\omega_a$ and $\omega_b$ produces the logarithmic form in Eq.~\eqref{eq:supp:rectangular-self-energy}, whose central region has a positive slope. Figure~S\ref{Fig_S10}(d) is a smoother, more realistic version: the logarithmic singularities in panel (e) are removed by the smooth frequency-dependence of self-energy, but the positive-slope region in $\Re \Sigma^R$ and the associated waterfall-like spectral feature remain. The pole expansion expresses the same mechanism in another form, because every Lorentzian bump in $-\Im f_j^R$ is accompanied by a positive slope in $\Re f_j^R$ for $|\omega-E_j|<\Gamma_j$. Whenever the combined positive slope approaches the condition in Eq.~\eqref{eq:supp:waterfall-condition}, the MDC-peak locus becomes nearly momentum independent over a finite energy interval and a waterfall-like spectral feature can emerge. This intrinsic mechanism is distinct from the extrinsic photoelectron energy-loss mechanism discussed in the main text, hence it should be viewed as a phenomenological reference rather than a material-specific identification of the origin of the observed waterfall-like spectral weight in LuFe$_6$Ge$_6$.

\bibliography{references}